\documentclass{article}
\usepackage{graphicx}
\usepackage{dcolumn}
\usepackage{bm}
\usepackage{color}
\usepackage{amsmath}

\begin{document}

\newcommand{\revise}[1]{\textcolor{red}{#1}}
\renewcommand{\vec}[1]{\mathbf{#1}}


%
%

\title{Using single-cell entropy to describe the dynamics of reprogramming and differentiation of induced pluripotent stem cells} 


\author{
Yusong Ye$^{1}$,
Zhuoqin Yang$^{1}$, 
Jinzhi Lei$^{2}$\footnote{Corresponding author, Email: jzlei@tiangong.edu.cn}
}

\maketitle

\begin{center}{
$^{1}$School of Mathematical Sciences, 
Beihang University, 
Beijing 100191, China
}
\end{center}

\begin{center}
{
$^{2}$
School of Mathematical Sciences,
Tiangong University,
Tianjin 30387, China
}
\end{center}


\begin{abstract}

Induced pluripotent stem cells (iPSCs) provide a great model to study the process of reprogramming and differentiation of stem cells. Single-cell RNA sequencing (scRNA-seq) enables us to investigate the reprogramming process at single-cell level. Here, we introduce single-cell entropy (scEntropy) as a macroscopic variable to quantify the cellular transcriptome from scRNA-seq data during reprogramming and differentiation of iPSCs. scEntropy measures the relative order parameter of genomic transcriptions at single cell level during the cell fate change process, which shows increasing during differentiation, and decreasing upon reprogramming. Moreover, based on the scEntropy dynamics, we construct a phenomenological stochastic differential equation model and the corresponding Fokker-Plank equation for cell state transitions during iPSC differentiation, which provide insights to infer cell fates changes and stem cell differentiation. This study is the first to introduce the novel concept of scEntropy to the biological process of iPSC, and suggests that the scEntropy can provide a suitable quantify to describe cell fate transition in differentiation and reprogramming of stem cells. 
\end{abstract}

\textit{single-cell RNA sequencing, single-cell entropy, induced pluripotent stem cell, stochastic dynamics, differentiation, reprogramming
}

\section{Introduction}
Induced pluripotent stem cells (iPSCs) are derived from skin or blood cells that have been reprogrammed back into an embryonic-like pluripotent state that enables the development of other types of differentiated cells. Reprogramming of iPSC is often induced by introducing genes important for maintaining the essential properties of embryonic stem cells (ESCs), and the genomic transcriptions changes during the process of reprogramming and further differentiation. However, this process is rather stochastic, and the molecular processes of cell fate changes remain unclear\cite{Stumpf:2017fa}. Recently, single-cell RNA sequencing (scRNA-seq) methods have allowed for the investigation of cellular transcriptome  at the level of individual levels\cite{takahashi2007induction,cacchiarelli2015integrative,gawad2016single}. The technique of scRNA-seq enables us to better study the dynamics of  reprogramming and differentiation of iPSCs, which can provide insightful informations on the process of cell reprogramming\cite{zhu2016integration,nicolae2010trait,gtex2017genetic}. Based on the microscopic state  of gene transcriptions provided by scRNA-seq, we can often determine the marker genes that guide the reprogramming process. Alternatively, a well defined macroscopic variable for the state of a cell is important for the understanding of the dynamics process.  

Recently, a novel concept of single-cell entropy (scEntropy) was proposed to measure the order of cellular transcriptome profile from scRNA-seq data\cite{liu2019single}. The  scEntropy of a cell is defined as the information entropy of the difference in transcriptions between the cell and a predefined reference cell, and provides a straightforward and parameter free macroscopic variable that can be used to quantify the process of early embryo development\cite{liu2019single}.  Here, we investigate whether scEntropy can be used to described the reprogramming and the differentiation process of iPSCs. We introduce the concept of scEntropy to quantify the states of individual cells along the reprogramming and differentiation processes, which reveal state changes during the biological processes. scEntropy can be served as a pseudo-time of the process, according to which we identify the genes that show expression correlated with scEntropy, and hence can be potent  marker genes of cell fate changes. We also constructed phenomenological stochastic differentiation equations for the plasticity process of stem cells. 

\section{Results}

\subsection{scEntropy to describe the process of differentiation and reprogramming}

The scEntropy was proposed to measure the order of cellular transcription from scRNA-seq data with respect to a reference level; larger entropy means lower order in the transcriptions\cite{liu2019single}. Given an $N\times M$ gene expression matrix with $N$ cells and $M$ genes, and the gene expression vector $\vec{r}$ of the reference cell. Let $\vec{x}_i\ (i=1, \cdots, N)$ the gene expression vector of the $i^\mathrm{th}$ cell. Calculation of the scEntropy of $\vec{x}_i$ with reference to $\vec{r}$, $S(\vec{x}_i | \vec{r})$, includes two steps\cite{liu2019single}: (1) calculate the difference between $\vec{x}_i$ and $\vec{r}$
$$\vec{y}_i = \vec{x}_i - \vec{r} = (y_{i1}, y_{i2},\cdots, y_{iM});$$
(2) the entropy $S(\vec{x}_i | \vec{r})$ is given by the information entropy of the signal sequence $\vec{y}_i$, \textit{i.e.},  
$$S(\vec{x}_i | \vec{r}) = - \int p_i(y) \ln p_i(y) \mathrm{d} y,$$
where $p_i(y)$ is the distribution density of the components $y_{ij}$ in $\vec{y}_i$. From the definition, the reference  cell $\vec{r}$ is a variable in defining the entropy $S(\vec{x}_i | \vec{r})$, which means the baseline transcriptome with the minimum entropy of zero.

To study the dynamics of reprogramming and differentiation of iPSCs, we can refer the state of iPSC as the reference cell in defining the scEntropy, the resulting scEntropy gives the relative information entropy of each cell with respect to the state of iPSC. 

To illustrate the application of scEntropy, we apply scEntropy to investigate the differentiation of iPSCs to cardiomyocytes. Time-series scRNA-seq data were obtained from 16 time points (the cells were sequenced every 24 hours for 16 days) in 19 human cell lines, totally 297 RNA samples were sequenced (GSE122380)\cite{strober2019dynamic}. To calculate the scEntropy, we take the average gene expressions of cells at D0 (the un-differentiated state) as the reference cell, so that scEntropy gives the relative transcription order with respect to the state of pluripotent stem cells. 

\begin{figure}[htbp]
\centering 
\includegraphics[width=0.6\linewidth]{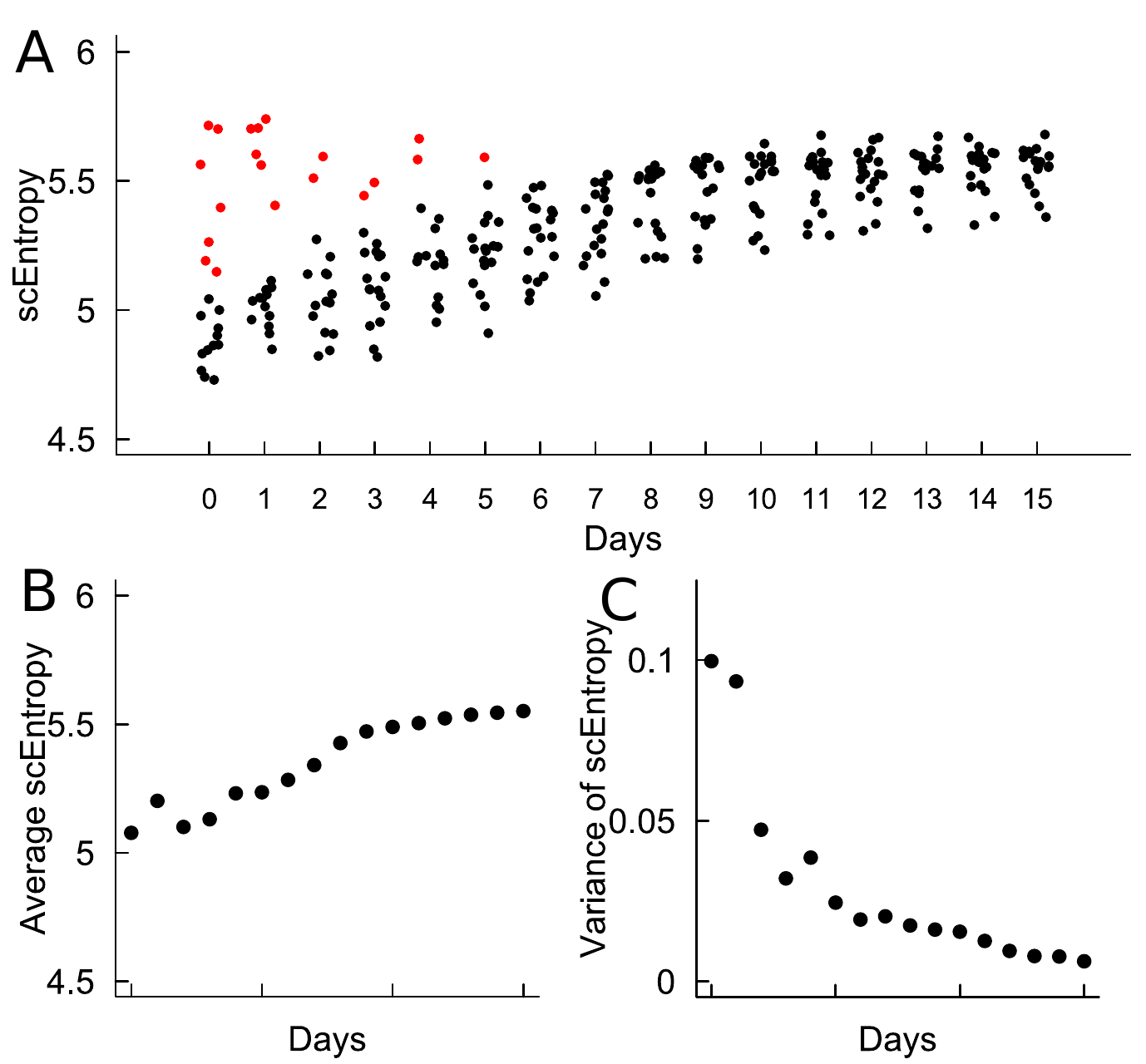}  
\caption{\textbf{scEntropy dynamics during the differentiation of iPSCs to cardiomyocytes.} \textbf{A}. scEntropies of 297 human iPSC cells sequenced every 24 hours for 16 days during the differentiation process. Black and red dots mark the two subgroup cells, respectively. See the text for details. \textbf{B}. Dynamics of the average scEntropies of cells sequenced at each day. \textbf{C}. Variance of the scEntropies of cells sequenced at each day.}
\label{fig1}
\end{figure}

The scEntropies of the sequenced cells at each day are shown at Fig. \ref{fig1}A. From Fig. \ref{fig1}A, there is an obvious tendency of increasing scEntropies along the differentiation process from D0 to D15. We also note the cell heterogeneity at D0. There are two group cells, group A cells show low level scEntropies with more stem-like cells (black dots in Fig. \ref{fig1}A), and group B cells show higher level scEntropies with less pluripotency (red dots in Fig. \ref{fig1}A). Experimentally, the reprogramming of iPSC procedure is not 100\% successful, the somatic cells are not reprogrammed synchronously, and some cells may fail to be induced to a pluripotency stem cells\cite{guo2019resolving,MacArthur:2013cv}, which correspond to the group B cells. Fig. \ref{fig1}A shows that group A cells shown increasing scEntropy during differentiation, and the two group cells emerge 8 days after the induction of differentiation.  

We further calculate the average and variance of the scEntropies of cells at each day (Fig. \ref {fig1}B-C). The average scEntropy obvious increases during differentiation, and the variance rapidly decreases from day 1 to day 2, along with the losing of heterogeneity. The cell-to-cell variance of scEntropies remain low at the later stages of differentiation, which may suggest the homogeneous dynamics of cell differentiation in the later stages (Fig. \ref {fig1}C). 

\begin{figure}[htbp]
	\centering 
	\includegraphics[width=0.6\linewidth]{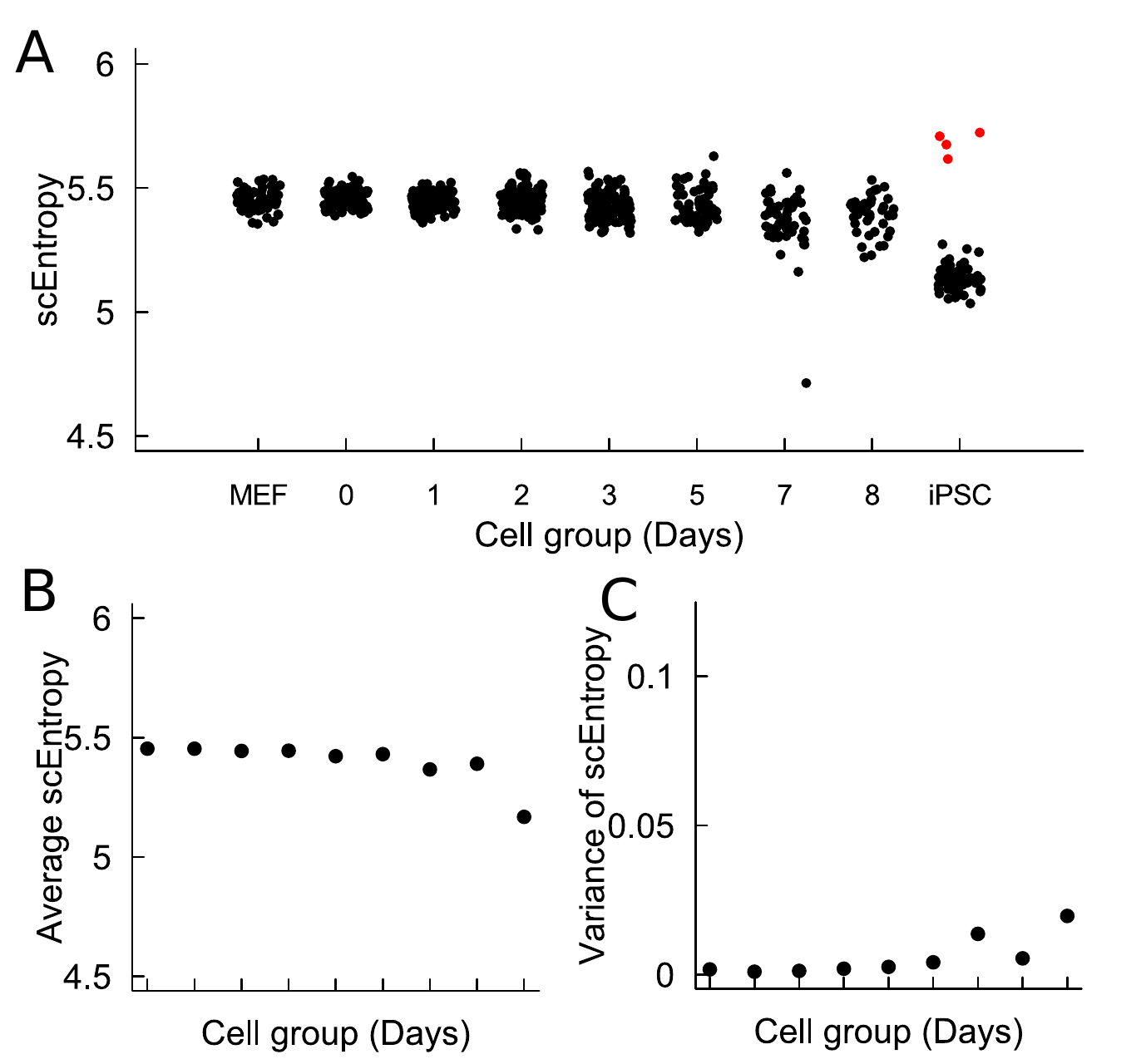}  
\caption{\textbf{scEntropy of cells during the process of reprogramming from MEF to iPSC}. \textbf{A}. scEntropies of cells in 9 time points during induced reprogramming. Black and red dots show the two subgroups cells, respectively. Referred to Fig. \ref{fig1} and the text for detials. \textbf{B}. Average of scEntropies of cells sequenced in each day.  \textbf{C}. Variance of scEntropies of cells sequenced in each day.}
\label{fig11}
\end{figure}

Next, we analyze the scEntropy of mouse cells during the reprogramming process from MEF to iPSC (GSE103221)\cite{guo2019resolving}. Totally 912 mouse cells were sequenced from 9 time points, scEntropies of the cells are calculated, and the reference cell is taken as the average gene expression vectors of all iPSCs. The scEntropy is nearly unchanged in 8 days after induction, and obviously decreases upon further reprogrammed into iPSC (Fig. \ref{fig11}A-B). We also note that a small fraction of cells (red dots in Fig. \ref{fig11}A) remain high scEntropy at the stage of iPSC (Fig. \ref{fig11}A), this may represent the cells that failed to be reprogrammed. The cell-to-cell variance of scEntropies of cells sequenced at the same day remain low along the reprogramming process (Fig. \ref{fig11}C). These results show changes in the order of cellular transcriptions during the reprogramming process. 

During cell reprogramming and differentiation, the gene expressions associated with cell pluripotency dynamically change in according with the cell types. The above applications show that scEntropy can quantify the processes, and show lower scEntropy for iPSCs comparing with the differentiated cells. We can consider scEntropy as an intrinsic state variable of the pluripotency of iPSCs, and hence changes of the scEntropy  can be served as a pseudo-time related to cell reprogramming/differentiation. Hereafter, we analyze the differentiation of iPSCs to cardiomyocytes to illuminate how scEntropy can help us to understand the biological processes of cell type changes.

\subsection{scEntropy as a pseudo-time of stem cell differentiation}

From Fig. \ref{fig1}, scEntropy increases with the differentiation process, and it measures the order of cellular transcription, which is an intrinsic state of a cell. Hence, we consider the scEntropy as a pseudo-time of each cell, which represents the change of intrinsic state of the cell along with differentiation.  

\begin{figure}[htbp]
	\centering 
	\includegraphics[width=0.9\linewidth]{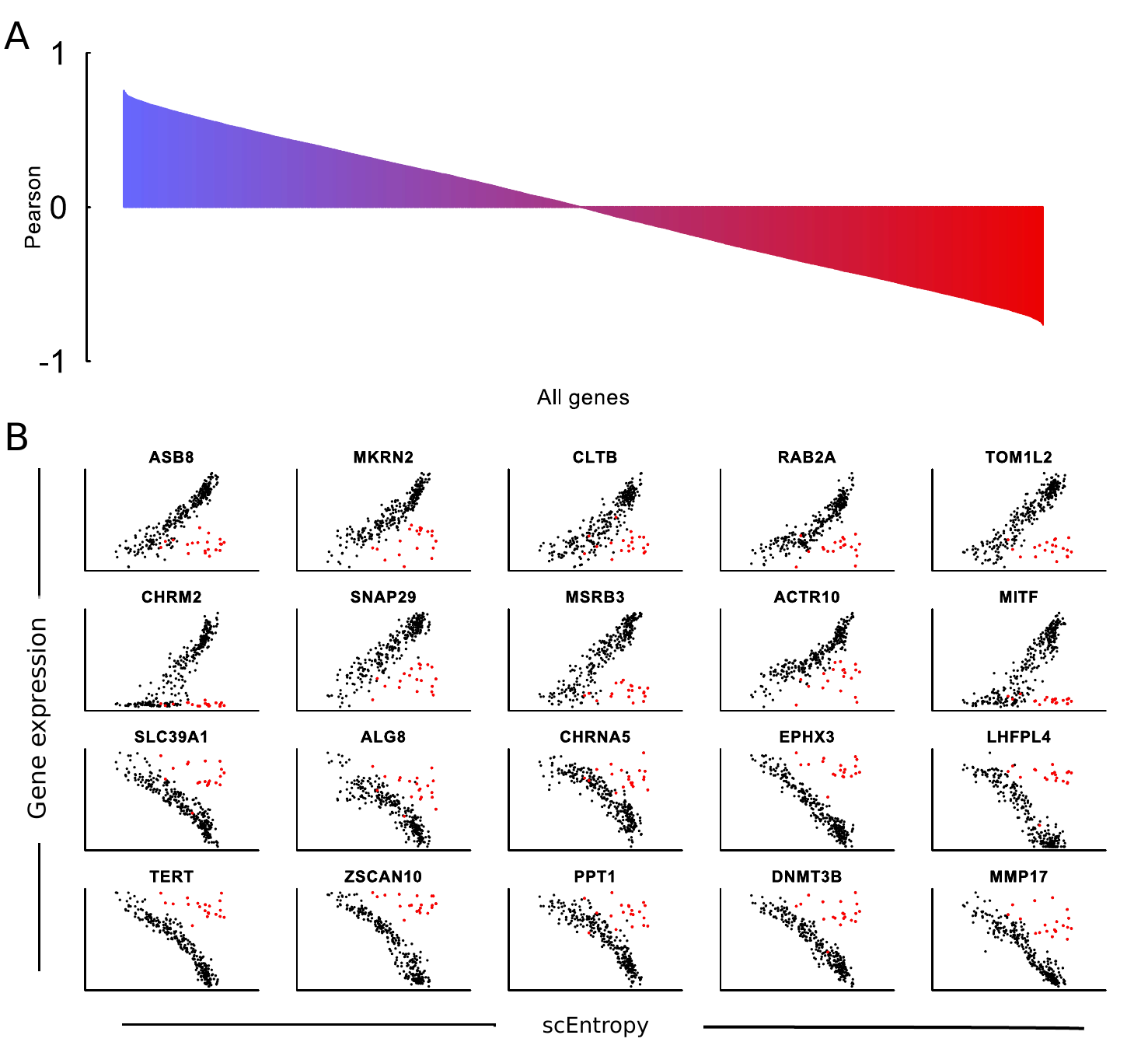}  
\caption{\textbf{Pearson correlation between gene expression and scEntropy}. \textbf{A}. Pearson correlation coefficient (PCC) of all 16319 genes. The genes are ordered according to the PCC value. \textbf{B}. Expressions of the top 10 positive correlation and top 10 negative correlation genes versus the scEntropy. Black and red dots represent the two subpopulation cells as in Fig. \ref{fig1}A.}
\label{fig2}
\end{figure}

Consider scEntropy as a pseudo-time, it is straightforward to study how the expressions of each gene vary over the differentiation process. Based on the scRNA-seq data of the differentiation of iPSCs to cardiomyocytes\cite{strober2019dynamic} (Fig. \ref{fig11}), we calculate the Pearson correlation coefficients of the expressions of each gene with the scEntropy. There are genes show high correlated (positive or negative ) with the scEntropy (Fig. \ref{fig2}A), these genes are potential marker genes of the differentiation process that show similar tendency of changes with the transcriptional order. We further analyze the top 10 positive correlated genes and the top 10 negative correlated genes (Fig. \ref{fig2}B). These genes include the pluripotent gene ZSCAN10, the DNA methyltransferase DNMT3B, and the telomerase reverse transcriptase (TERT) which are closely related to the cell division process. From Fig. \ref{fig2}B, the two subgroup cells in Fig .\ref{fig1}A (black and red, respectively) show different dependence in gene expression with respect to change in the scEntropy. The group A cells (black dots) show increasing (positive correlated genes) or decreasing (negative correlated genes) with the scEntropy. In group B cells (red dots), however, the expression of these genes show independent with changes of the scEntropy. In these cells, expressions of the above genes shown a weak correlation with the scEntropy and are different from the other differentiated cells.  For example, the expression level of the gene CHRM2 shown nearly unchanged with the increasing of the scEntropy. These results reveal the existence of two type cells at D0, the iPSCs that can differentiate to cardiomyocytes, and the un-reprogrammed cells that remain phenotypically unchanged during the induction of differentiation.  
  
 \begin{figure}[htbp]
	\centering 
	\includegraphics[width=1.0\linewidth]{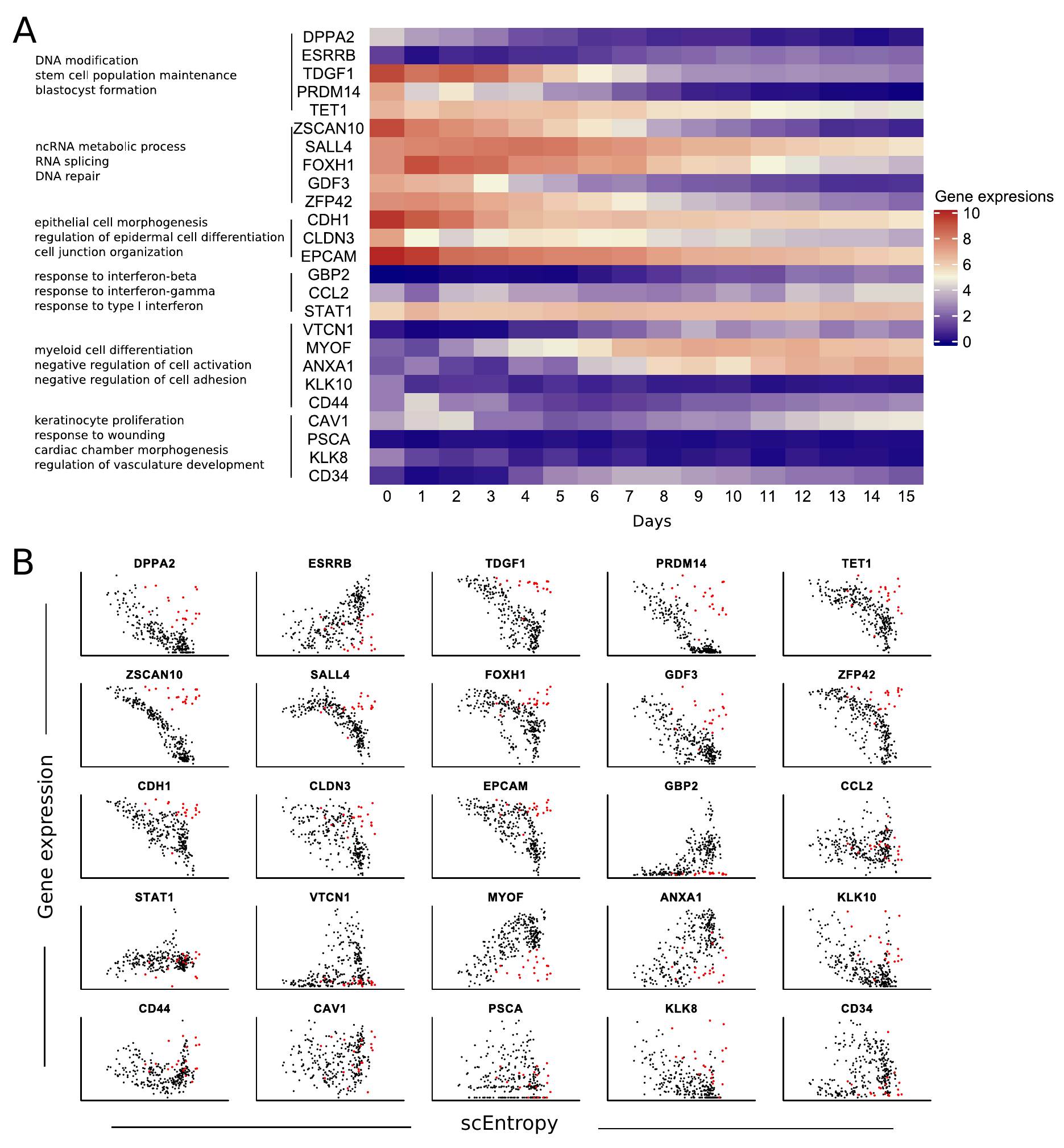}  
\caption{\textbf{Marker genes during the differentiation process}. \textbf{A}. Average gene expressions of 25 marker genes along the differentiation process. \textbf{B}. Expressions of the 25 genes versus scEntropy.}
\label{figheat}
\end{figure}

To further analyze the correlation between scEntropy and gene expressions, we identify 25 genes through GO enrichment that are associated with functions related to cell pluripotency, differentiation, and DNA maintenance, \textit{etc}. Expressions of these genes along the differentiation process are shown as the heat map in Fig. \ref{figheat}A. Dynamics of these gene expressions with the increasing of scEntropy are shown in Fig. \ref{figheat}B.  We show that some genes obviously decrease with scEntropy, such as the pluripotent genes ZSCAN10 and TET1. Expressions of TET1 decreases along the differentiation. TET1 usually up-regulate transcription by RNA polymerase and  promotes DNA demethylation process in differentiation. ZSCAN10 is another interesting gene that negative correlated with the scEntropy and decrease during differentiation. ZSCAN10 is known to function with somatic-stem cell population maintenance and transcription factor activity. The non-reprogramming relevant gene PSCA, KIK8, and CD34 show no correlation with the scEntropy, and nearly unchanged during differentiation. CD44 and SALL4 provide another scenario. These genes are enrich in the pathways of DNA binding and stem cell population maintenance. CD44 obvious increase when the scEntropy is large, and SALL4 decrease with the increasing of scEntropy. These results show that when we consider scEntropy as a pseudo-time of the differentiation process, the correlation between gene expressions and the scEntropy can provide informations on how gene expression changes with the intrinsic state of cells. 

\subsection{scEntropy dynamics and the transition of cell fates}

\begin{figure}[htbp]
	\centering 
	\includegraphics[width=0.7\linewidth]{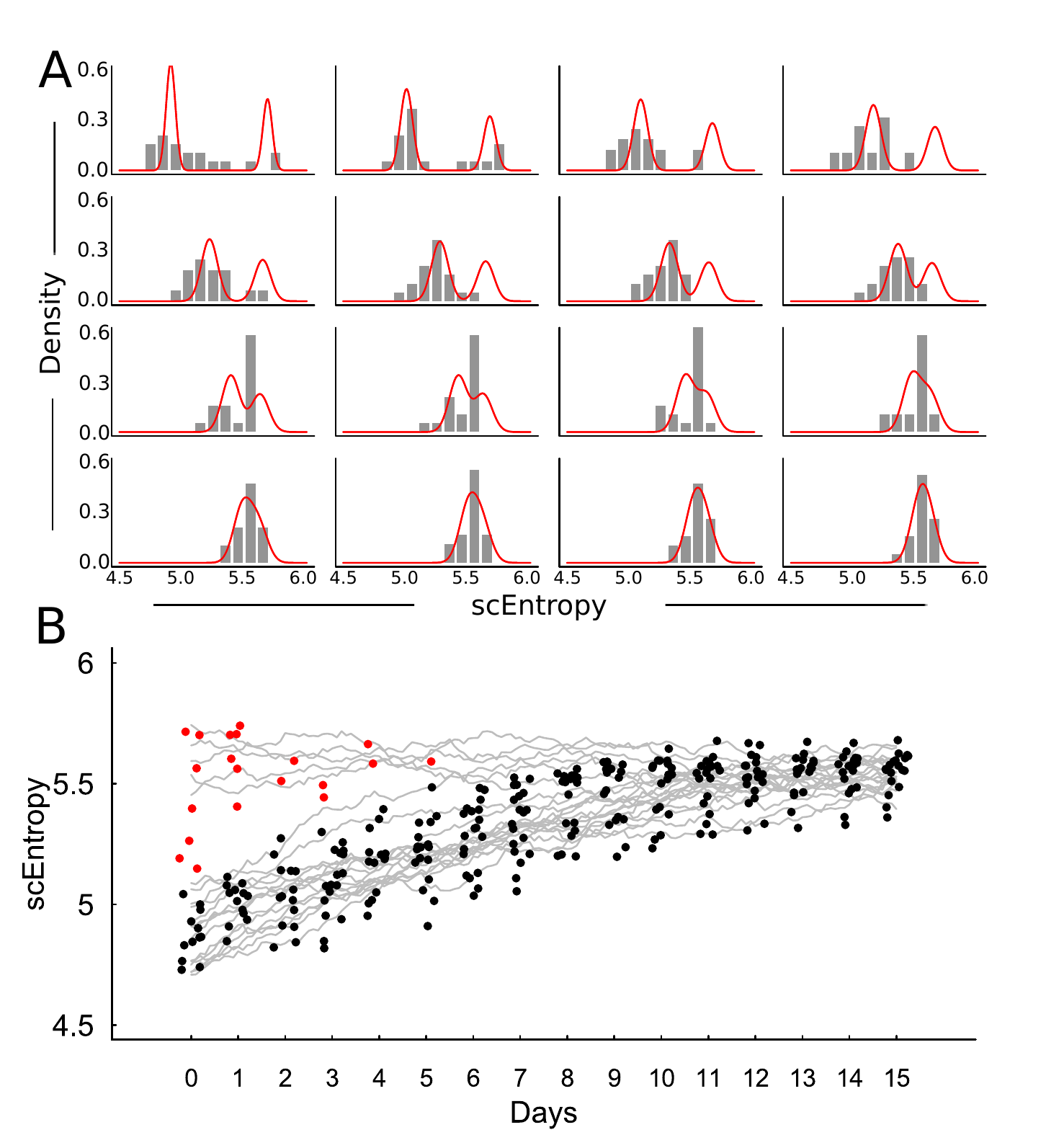}  
\caption{\textbf{Transition of cell fates during the differentiation of iPSCs to cardiomyocytes}. \textbf{A}. The scEntropy distribution (histogram) during the differentiation of iPSCs to cardiomyocytes. Red lines show the theoretical distribution  given by \eqref{eqdist} from the solution the Fokker-Planck equation \eqref{eqfpe}. \textbf{B}. Simulated scEntropy dynamics based on the stochastic differential equation \eqref{eqsde}. Dots are scEntropy from the scRNA-seq data (referred to Fig. \ref{fig11}). Here, the parameters are $k=0.15\ \mathrm{day}^{-1}$, $B=0.04$, $x^*=5.6$, and the initial conditions are taken either low or high level scEntropies. }
\label{fig3}
\end{figure}

To further investigate the dynamic process of cell date changes, we analyze the distribution of scEntropies of cells at each day during the differentiation of iPSCs to cardiomyocytes (Fig. \ref{fig3}A). The distribution has two peaks at D0, corresponding to pluripotent cells with low entropy, and non-reprogrammed cells with high entropy, respectively. After the induction of differentiation, the entropy of pluripotent cells increase with the differentiation process, and the number of non-reprogrammed cells decreases. Finally, there is only one peak in the distribution from D8. The increasing of the scEntropy of pluripotent cells during differentiation suggests the decreasing of transcriptional order in the differentiation of iPSCs to cardiomyocytes. 
 
From the above analysis, the distribution of scEntropy evolves to a unimodal distribution in 16 days, which suggest that the cell scEntropy converge to a stationary state of a stable scEntropy. Hence, we can describe the differentiation process of a single cell through the dynamics of its scEntropy, which is formulated as a stochastic process through a stochastic differentiation equation. Let $X_t$ represents the scEntropy of a cell at time $t$, and $x^*$ the average scEntropy at the  differentiated state, we introduce the following Ornstein-Uhlenbeck process to model the phenomenological dynamics of the scEntropy
\begin{equation}
\textrm{d}{X(t)}=-k(X-x^*) \textrm{d}t + B \textrm{d}W_t,
\label{eqsde}
\end{equation}
here $k$ is a parameter describing dissipation velocity, $B$ is the fluctuation parameter, and $W_t$ means the Weiner process. Given the parameters and initial condition, a sample solution of \eqref{eqsde} gives a possible trajectory of scEntropy of a single cell during differentiation. Fig. \ref{fig3}B shows the trajectories obtained from \eqref{eqsde}.  Currently, we can only sequence a cell once, and it is impossible to track the evolution of transcriptome for a single-cell. Here, the equation \eqref{eqsde} provides a conceptual  description of the transcriptional states of a cell during a process of cell fate decision.  

From the stochastic differential equation, it is straightforward to obtain the associated Fokker-Planck equation
\begin{equation}
\frac{\partial f(x,t)}{\partial t}=-\frac{\partial}{\partial x}(-k(x-x^*) f(x,t))+
\frac{B^2}{2}\frac{\partial^2}{\partial x^2}( f(x,t))
\label{eqfpe}
\end{equation}
Here, $f(x, t)  = P\{ X(t) = x\}$ means the probability of a cell to have scEntropy $x$ at time $t$. In  particular, given the initial state $X(0) = x_0$, the transition probability $P\{X(t) = x | X(0) = x_0)$ can be obtained by solving the equation \eqref{eqfpe} with initial condition $f(x, 0) = \delta(x-x_0)$, which is given by a Gaussian distribution with mean $\varphi(t; x_0) = x^* + (x_0-x^*) e^{-k \cdot t}$ and variance $\sigma^2(t)=\frac{B^2}{2 k}(1-e^{-2k t})$:
\begin{equation}
P\{X(t)= x | X(0) = x_0\}=\frac{1}{ \sqrt{2 \pi} \sigma(t)} e^{-\frac{(x - \varphi(t; x_0))^2}{2 \sigma^2(t)}}
\label{eqsolu}
\end{equation}
We fit \eqref{eqsolu} with the daily distributions shown by the histograms in Fig. \ref{fig3}A. In fitting the data, we take the initial values $x_0 = 4.8$ with probability $0.6$ and $x_0  = 5.7$ with probability $0.4$, the distributions are fitted with (here $t=0$ corresponds to day $-1$ before the onset of cell differentiation)
\begin{equation}
\label{eqdist}
f(x,t) = \frac{1}{ \sqrt{2 \pi} \sigma(t)} \left(0.6 e^{-\frac{(x - \varphi(t; 4.8))^2}{2 \sigma^2(t)}} + 0.4 e^{-\frac{(x - \varphi(t; 5.7))^2}{2 \sigma^2(t)}}\right)
\end{equation}
The obtained parameters are $k = 0.15 \mathrm{day}^{-1}, B=0.04$, and the theoretical distributions are shown by red lines in Fig. \ref{fig3}A. 

\section{Discussion}

The novel concept of scEntropy has been defined to measure the macroscopic transcription order of individual  cells based on scRNA-seq\cite{liu2019single}. The concept of scEntropy is valuable in describing the process of early embryo development, as well as the classification between normal and malignant cells in different types of cancers\cite{liu2019single}. The current study is the first to introduce scEntropy as a macroscopic quantity of the transcription state of cells to describe the process iPSCs reprogramming and differentiation. The scEntropy can be a pseudo-time of the differentiation/reprogramming process and how decreasing with the increasing of cell pluripotency.  The proposed scEntropy is an intrinsic quantify of the cell transcriptional state, and hence genes that show gene expressions correlated with the scEntropy can be essential for the process of cell fate decision. We can also consider the dynamics of scEntropy of an individual cell during cell fate transition as a stochastic process, which can be modeled through Langevin equations. The conceptual model  of Langevin equation provide insights on the dynamics of cell state changes during the differentiation/reprogramming of iPSCs. Based on the stochastic dynamics of the scEntropy, the epigenetic landscape of cell differentiation/reprogramming can be described by the related Fokker-Plank equation. Our results shown the cell-to-cell variability of scEntropy during the differentiation process. Similarly, heterogeneity of gene expression measured by the Shannon entropy also increase in the first few hours in the differentiation process of chicken erythroid progenitors\cite{Richard:2016ba} and in the differentiation from pluripotent stem cells to neuronal state\cite{Stumpf:2017fa}. Both entropies from cell- and gene- based show increase during the differentiation process, which suggest the features of variability and stochasticity in stem cell differentiation.

The scEntropy was proposed to measure the intrinsic order of the cellular transcriptome with respect to a predefined reference cell. The definition of scEntropy includes no external parameters, and hence can provide natural informations of a cell. It is essential to quantify the changes of intrinsic cellular transcriptome  during differentiation/reprogramming for our understanding of the biological process of cell fate decision. iPSCs provide a controllable system to study the cell fate changes at individual cell level. Single-cell sequencing techniques enable us to examine the microscopic states of individual cells. However, there are two major limitations when we analyze the single-cell sequencing data: each cell can only be sequenced once and hence not able to track a cell dynamics; and the sequencing data are usually very high dimensional, it is usually difficult to find the low dimensional features to characterize the cell.  Hence, for a better understanding of the cell type transition process, it is important to quantify the cell types through the intrinsic state of a cell. Our study shows that scEntropy can be one of the potent variable for the intrinsic state of cellular transcriptome, and can be used to describe the single cell dynamics of epigenetic landscape through the stochastic dynamical models. The current approach can be extended to explore more detail dynamics of various biological processes of cell type transitions, such as early embryo development, cancer cell plasticity, cell differentiation. 

\section{Acknowledgments}
This work is supported by the National Natural Science Foundation of China (11831015, and 11872084).


\end{document}